# A low temperature disordered phase of α-Pb/Ge(111)

Jiandong Guo[1], Junren Shi[2], and E. W. Plummer[1,2]

[1]*Department of Physics and Astronomy, University of Tennessee, Knoxville, TN 37996*
[2]*Condensed-Matter Science Division, Oak Ridge National Laboratory, Oak Ridge, TN 37831*

A new structural phase transition has been observed at low temperatures for the one third of a monolayer (α phase) of Pb on Ge(111) using a variable-temperature scanning tunneling microscope. The well-known ($\sqrt{3} \times \sqrt{3}$)R30° to (3×3) transition is accompanied by a new structural phase transition from (3×3) to a disordered phase at ~76 K. The formation of the new disordered phase is incompatible with the belief that the (3×3) phase is the ground state. The mechanism of the phase transition in this triangular lattice can be mapped onto antiferromagnetic spin Ising model, with the geometric frustration.

Two-dimensional (2D) metallic films on semiconducting substrates are inherently unstable. The metallic electrons tend to drive surface reconstruction so that the energy of the electron subsystem can be lowered by forming an insulating state. In contrast, the coupling of the thin metallic film to the substrate may prefer no reconstruction. The equilibrium structure is essentially a compromise between these two driving forces. Such a competition is responsible for the great varieties of structural phase transitions observed in metal-adsorbed semiconductor surfaces.

A system that has received considerable attention is the symmetry-lowering structural phase transition observed in the one-third of a monolayer of Pb adatoms atop of the Ge(111) surface (α-phase Pb/Ge interface) [1]. At room temperature (RT), the Pb atoms are arranged to a ($\sqrt{3} \times \sqrt{3}$)R30° structure (referred to as ($\sqrt{3} \times \sqrt{3}$) in the following), but become distorted forming a (3×3) symmetry at lower temperatures. The (3×3) results from an ordered array of *one up – two down* configuration of the Pb atoms [1]. This phase transition in Pb/Ge and in the isoelectronic Sn/Ge interface [2] has been a subject of extensive experimental and theoretical investigation [3-10]. There is still discussion concerning the origin of this transition, especially when the role of defects is included [11-14]. A recent theoretical paper [15] made an interesting prediction. When the competition between what the film and the substrate want to do is right a low temperature (LT) "glassy" phase can develop.

This letter reports a variable-temperature scanning tunneling microscope (STM) study on the Pb/Ge interface at reduced temperatures. In addition to the ($\sqrt{3} \times \sqrt{3}$) to (3×3) transition, a transition from (3×3) to a disordered phase is observed upon cooling the sample to below 76 K. In this new phase, the Pb adatoms distort with no apparent long-range order (LRO), resulting in an irregular "kinked-line" structure. Although defects blur both phase transitions (as they did in the Sn/Ge system [13, 16]), it is demonstrated by direct STM imaging that the observed disordered phase is intrinsic to the Pb/Ge interface.

The experiments were performed in an Omicron ultra high vacuum variable-temperature STM. The Ge(111) substrate was cleaned *in situ* by repeated sputtering and annealing cycles until nice c(2×8) surface reconstruction was obtained. Pb was deposited from a commercial Knudsen cell (K-Cell) for 45 seconds. During deposition the temperature was stabilized at 550°C in the K-Cell and at 100°C on the substrate, respectively. Finally, the sample was annealed at 360°C for 20 minutes. Well-ordered interface with exact Pb coverage at one-third of a monolayer was verified by the sharp ($\sqrt{3} \times \sqrt{3}$) LEED patterns that appeared as soon as the preparation finished. LT LEED was also performed by cooling the sample with liquid nitrogen. With an onset at 250 K, extra diffraction spots corresponding to (3×3) symmetry became visible and intensified at decreased temperature. This structural transition is gradual and reversible. RT STM images revealed the existence of Ge substitutional defects and negligible amount of vacancies on the ($\sqrt{3} \times \sqrt{3}$) reconstructed surface. By optimizing the post-evaporation annealing temperature and time, the defect density was minimized to about 10% with average distance of about 2 nm between nearest neighboring defects. The sample was then cooled on the STM stage by using a continuous flow cryostat. Temperature at the sample surface was controlled and stabilized in the range between RT and 40 K. All the STM images presented in this letter were obtained with the tip bias at +1.5 V (filled state) and feedback current at 3 nA. Since the filled-state images are insensitive to the tip bias, it is assumed that these high-bias images mirror the structural information of the Pb/Ge interface directly.

Figure 1(a) - (c) summarize the STM observation of the Pb/Ge interface at different temperatures. The RT image (a) reveals the ($\sqrt{3} \times \sqrt{3}$) structure with the dark spots being the substitutional Ge atoms [11]. Upon cooling, the Pb atoms display an ordered vertical distortion producing (3×3) symmetry at about 110 K (Fig. 1(b)). The STM image represents the *one up - two down* configuration (one bright - two dark in the filled state image). The (3×3) phase is stable between 110 K and 80 K with a size of ~100 nm$^2$. As the temperature is decreased the (3×3) domains start to shrink

and the domain walls dissolve. In Fig. 1 (c), the Pb atoms show a disordered structure at 41 K in which the Pb atoms are at randomly distributed height with no apparent LRO [17]. Unlike the (3×3) phase, no preferred positions (perpendicular distortion) can be identified in the STM images. The Pb atoms exhibit a continuous distribution of heights. All transitions from the ($\sqrt{3} \times \sqrt{3}$) to the (3×3) and from the (3×3) to the disordered phase appear to be reversible.

These two phase transitions can be clearly seen in the Fourier transformations (FT) of the real space STM images (insets in fig. 1(a)-(c)). The ($\sqrt{3} \times \sqrt{3}$) hexagon in the FTs appears in all three phases (a)-(c). This substrate holds the Pb atoms horizontal position in this basic triangular structure. The hexagon (3×3) spots in the FT are weak and fuzzy for the "flat" ($\sqrt{3} \times \sqrt{3}$) surface at RT, turning sharp for the surface buckled with (3×3) LRO, and become fuzzy again for the randomly corrugated surface at 41 K. The presence of defects in the interface results in the enhancement of the (3×3) symmetry in the FTs at all temperatures, which will be discussed subsequently.

To focus attention on the vertical distortions in the STM images the ($\sqrt{3} \times \sqrt{3}$) spots in the FT are removed. The transform back to real space is shown in Fig. 1 (d) – (f). At RT (Fig. 1(d)), all the Pb atoms are equivalent except small distortions pinned around defects. This local distortion has the same characteristics as reported for the Sn/Ge system [11-13,16]. At 90 K in (e), large (3×3) domains and sharp domain walls are present, indicating the new LRO is fully developed. In the disordered phase at 41 K, there is no apparent LRO and the configuration of the distorted Pb atoms can be best described as "kinked-lines", which is indicated by the gray lines in Fig. 1 (f).

Two order parameters are defined to quantify these two transitions. First, the atomic roughness ($\Delta Z$), which is defined as the standard deviation of adatom heights seen by the STM, distinguishes the "flat" ($\sqrt{3} \times \sqrt{3}$) phase and the distorted phases at LT. As shown in the upper panel of Fig. 2, the order parameter $\Delta Z(T)$ increases continuously upon cooling followed by a saturation at $T_1$. The abrupt change of the order parameter expected in a phase transition is absent due to the presence of defects that induce local distortions in the lattice [12,13]. The transition temperature $T_1$ is estimated from the kink point of the temperature dependence of $\Delta Z$, and yields $T_1 \sim 112$ K. The magnitude of $\Delta Z$ is 3 times larger than that determined by the surface x-ray diffraction (SXRD) analysis [5], indicating that the charge redistribution is more dramatic than the lattice distortion.

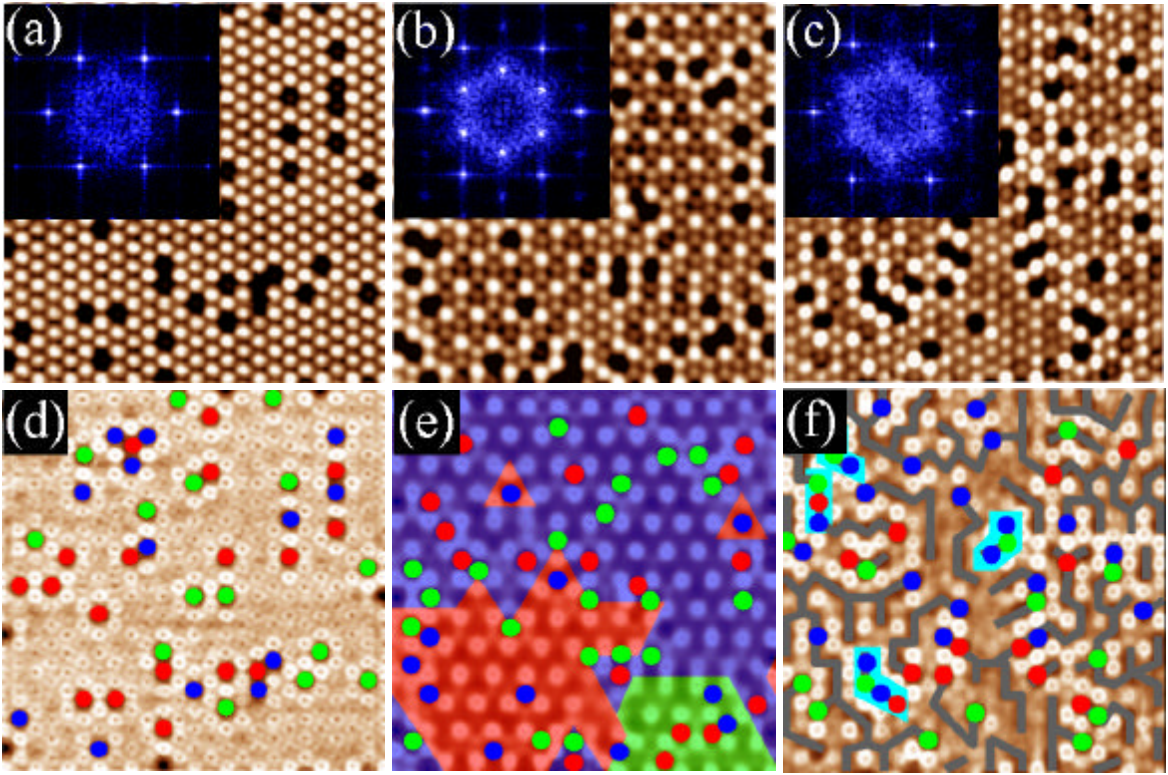

FIG. 1    STM images (14×14 nm$^2$) of the Pb/Ge interface at (a) RT, (b) 90 K, and (c) 41 K. STM images presented were obtained with the tip bias at +1.5 V and feedback current at 3 nA. The FTs are shown as insets of the corresponding real space images. (d)-(f) show the perpendicular distortion by filtering out the ($\sqrt{3} \times \sqrt{3}$) symmetry from images (a)-(c), respectively. Ge substitutional defects are marked with red, green, or blue circles depending on their coordination relative to the three nonequivalent sites in a (3×3) lattice. Three or more neighboring defects are marked in light blue in (f). In (e), the (3×3) domains are highlighted in colors of the "up" sites. The adatoms distorted downwards are connected with gray lines in (f).

The second order parameter is $I_3/I_{\sqrt{3}}$, defined as the intensity ratio between the (3×3) and ($\sqrt{3}\times\sqrt{3}$) spots in the FT of the real space image. This order parameter will distinguish the (3×3) phase from the disordered phase at LT. As shown in the lower panel of Fig. 2, $I_3/I_{\sqrt{3}}$ decreases upon cooling below $T_2 \sim 76$ K, signifying the disappearing of the (3×3) LRO and the onset of the disordered phase. $I_3/I_{\sqrt{3}}$ also shows the same $T_1$ that was seen in $\Delta Z(T)$.

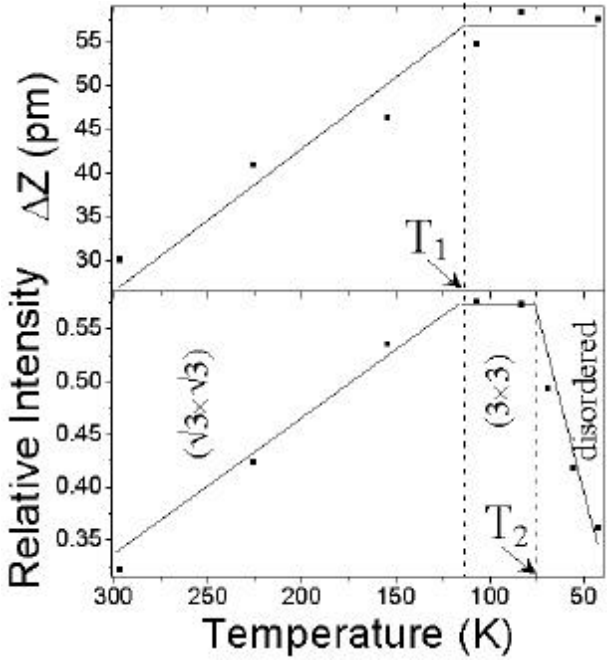

FIG. 2 Temperature dependence of the two order parameters of Pb/Ge interface $\Delta Z(T)$ and $I_3/I_{\sqrt{3}}$ (T). The height Z was obtained with the tip bias at +1.5 V and feedback current at 3 nA.

It has been documented that the ($\sqrt{3}\times\sqrt{3}$) to (3×3) transition, especially for the Sn/Ge system, is influenced by defects [11-14,16]. A quick survey of our STM images reveals that defects are behaving similarly for the Pb/Ge system. Fig. 1(d) shows a honeycomb structure developing around the Ge defects [12,13,16]. The range of this defect induced density wave is temperature dependent [12,16] and is responsible for the high temperature tail seen in both order parameters $\Delta Z$ and $I_3/I_{\sqrt{3}}$, blurring a clear definition of $T_1$. When the system enters into the global (3×3) phase below $T_1$, the defects develop a correlation between themselves in such a way that almost all defects are aligned to the "down" sites (the charge-minimum sites) of the (3×3) domains, as previously reported as the "defect-ordering transition" in the Sn/Ge interface [16]. Such a correlation between defects can be clearly seen in Fig. 1 (e), which shows that all the defects are either isolated or forming pairs, compatible to the (3×3) LRO with *one up – two down* configuration within each domain (red, green or blue). Fig. 1(f) shows that there is a defect induced local distortion at 40 K but it is quite different. First, the defects seem lost their alignment within domains as seen in Fig. 1 (e), probably because the domain size is very small. Secondly, it is clear that there is a distinct configuration in which three or more defects may sit side by side and are arranged to "lines" or "kinked-lines", as highlighted in Fig. 1 (f). Such a change of the defect distribution indicates that the defects have again moved to accommodate the configuration in the disordered phase.

It is important to clarify the issue whether the observed disordered phase is an intrinsic behavior of the Pb/Ge interface, or an effect determined by the defects. These measurements indicate that the defects move on the surface to accommodate to the lowest energy configuration for each phase: random in the ($\sqrt{3}\times\sqrt{3}$), ordered within a domain for the (3×3), and in a "line-kinked" configuration for the distorted phase.

The intrinsic change of the Pb/Ge interface structure becomes clearer by observing different responses of the underlying lattice to defects in the (3×3) phase and in the disordered phase. In Fig. 3, we show two areas in the (3×3) phase at 90 K (left) and the disordered phase at 40 K (right), respectively. Both areas are in the same size and have the same number of defects. In the (3×3) phase, a large (3×3) single domain is pinned around the 8 defects with distribution compatible to the *one up – two down* (3×3) symmetry. The local distortion around defects is in harmony with the global lattice structure as evident in the left panel of Fig. 3. On the other hand, the right panel of Fig. 3 shows that the kinked-line structure is developed at LT in the area with 8 defects even though a single (3×3) domain could accommodate all of them. This comparison clearly demonstrates that the disordered structure is intrinsically a new phase determined by temperature only but not defects. It is important to point out that such a defect distribution is not representative at LT and corresponding to a small portion of defects not fully relax to equilibrium sites in the interface due to the finite cooling rate in the experiments. These unrelaxed defects result in a frustrated local configuration around them in the disordered phase, while in the (3×3) phase they are responsible for the domain walls (see the right panel of Fig. 3 and Fig. 1 e).

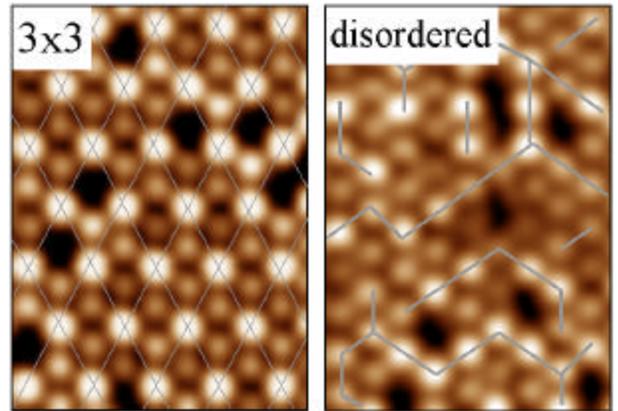

FIG. 3 STM images of a (3×3) single-domain at 90 K and an area in the disordered phase at 40 K with the same defect distribution with respect to the virtual (3×3) lattice. The gray grids in the left panel mark the charge-maximum sites of the (3×3) lattice. The kinked lines structure is indicated in the right panel by the gray lines connecting the adatoms distorted upwards.

Existence of the disordered phase is contradictory to the traditional belief that the ordered phase is favored at LT. Among all the theories dedicated to understand the ($\sqrt{3}\times\sqrt{3}$) to (3×3) phase transition at Sn/Ge or Pb/Ge interface, only the one proposed recently by Shi *et al.* [15] predicted such a transition from the (3×3) phase to the disordered phase. In that theory, the driving force of the transitions is the electron-mediated adatom-adatom interaction and the "glassy" phase originates from the geometric frustration associated with the triangular arrangement of adatoms. The electron-mediated adatom-adatom interaction arises because the motion of adatoms is coupled with the motion of electrons (electron-phonon coupling) in a metallic interface. In the Pb/Ge and isoelectronic systems, moving down an adatom repels electrons from the lattice site and the excess electrons are absorbed by the neighboring adatoms, rendering them to move up. This results in a repulsive interaction between the neighboring adatoms. When such a repulsive adatom-adatom interaction overwhelms the local stress imposed by substrate that keeps the Pb atoms undistorted, a structural phase transition occurs. The transition is accompanied by the surface charge redistribution with more electrons at "up" sites and less electrons at "down" sites, which is consistent with our observation that the distortion observed in STM is much larger than that determined from SXRD [5].

Depending on the relative strength of the electron-mediated interaction to the local stress, the new phase stabilized at LT could be the (3×3) phase or the "kinked line" glassy phase. The glassy phase occurs when the electron-mediated adatom-adatom interaction is much stronger than the local stress. In this limit, the system can be mapped to an antiferromagnetic spin Ising model. Simplifying the perpendicular distortion of the adatoms into "up" or "down" relatively to their average height without counting the distortion quantitatively, the structural up (down) adatoms are corresponding to the spin-up (spin-down) while the repulsive electron-mediated adatom-adatom interaction are corresponding to the antiferromagnetic coupling between spinons. It is well known that an antiferromagnetic Ising system shows frustration in a triangular ($\sqrt{3}\times\sqrt{3}$) lattice with kinked-line arrangement of spins [18], exactly what we have observed in the corresponding Pb/Ge thin film system at LT.

Although the theory provides a consistent picture for the experimental observations, we need to point out some discrepancies. (1) The experimentally determined transition temperature to the disordered phase ($T_2$) is much higher than the theoretical prediction presented in Ref. 15. This may be due to the over-simplified nature of the model employed in the theory. The presence of defects may also enhance the transition temperature to the glassy phase. (2)

A recent LEED study on Si(001) surface [19] also observed the absence of the LRO at LT. The geometric frustration cannot be the cause in this case because the surface atoms are arranged into a rectangular lattice. Although more experimental studies are required before concluding whether the transition in Si(001) has the same nature as that in Pb/Ge interface, a question has arisen that if there is a mechanism more general than the simple geometric frustration and applicable to other structures.

In summary, a 2D disordered phase is observed in Pb/Ge interface below 70 K with randomly distorted structure. Two order parameters are introduced to characterize quantitatively the ($\sqrt{3}\times\sqrt{3}$) to (3×3) transition at about 112 K and (3×3) to disordered phase transition at about 76 K. Both transitions are demonstrated to be intrinsic to the Pb/Ge interface although blurred by the presence of defects. The experimental observations are consistent with a previous theory prediction that the system may enter into the glassy phase due to the geometric frustration of the ($\sqrt{3}\times\sqrt{3}$) structure and the repulsive electron-mediated interaction between adatoms.

The authors are grateful to the discussions with A. V. Melechko and Jiandi Zhang. This work is funded by NSF DMR-0105232. Oak Ridge National Laboratory is managed by UT-Battelle, LLC, for the U. S. Department of Energy under Contract No. DE-AC05-00OR22725.

[1] J. M. Carpinelli *et al.*, Nature **381**, 398 (1996).
[2] See a recent review, T-C Chiang *et al.*, J. Phys: Condens. Matter **14**, R1 (2002).
[3] G. Le Lay *et al.*, Appl. Surf. Sci. **123/124**, 440 (1998).
[4] R. I. G. Uhrberg *et al.*, Phys. Rev. Lett. **81**, 2108 (1998).
[5] O. Bunk, *et al.*, Phys. Rev. Lett. **83**, 2226 (1999).
[6] J. Zhang *et al.*, Phys. Rev. B **60**, 2860 (1999).
[7] R. Pérez *et al.*, Phys. Rev. Lett. **86**, 4891 (2001).
[8] José Ortega *et al.*, J. Phys.: Condens. Matter **14**, 5979 (2002).
[9] G. Santoro *et al.*, Comp. Mat. Sci. **20**, 343 (2001).
[10] S. de Gironcoli *et al.*, Surf. Sci. **454-456**, 172 (2000).
[11] H. H. Weitering *et al.*, Science **285**, 2107 (1999).
[12] A. V. Melechko *et al.*, Phys. Rev. Lett. **83**, 999 (1999).
[13] L. Petersen *et al.*, Phys. Rev. B **65**, 020101 (2002).
[14] T. E. Kidd *et al.*, Phys. Rev. Lett. **85**, 3684 (2000).
[15] J. Shi *et al.*, Phys. Rev. Lett. **91**, 76103 (2003).
[16] A. V. Melechko *et al.*, Phys. Rev. B **61**, 2235 (2000).
[17] As evident of the absence of LRO, no sharp domain walls are visualized below T$_2$. But in some rare cases, small (3×3) areas with the diameter up to 3 nm are observed as low as ~60 K which is compatible with Ref. 1. As discussed subsequently in text, the slightly higher defect density might further increase the transition temperature to the disordered phase in our sample.
[18] K. H. Fischer and J. A. Hertz, *Spin Glasses*, Cambridge University Press (1991).
[19] M. Matsumoto *et al.*, Phys. Rev. Lett. **90**, 106103 (2003).

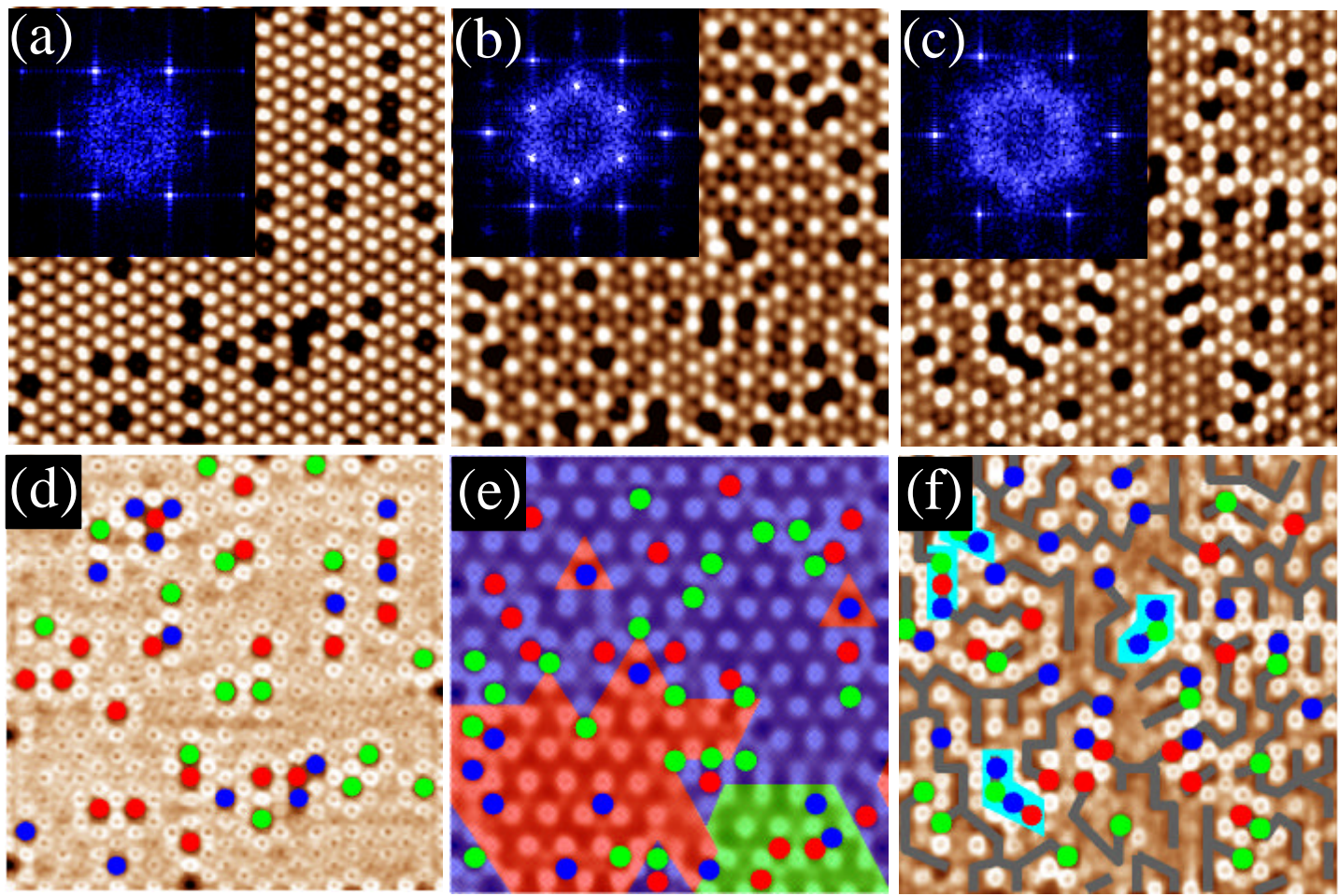

FIG. 1 STM images (14×14 nm$^2$) of the Pb/Ge interface at (a) RT, (b) 90 K, and (c) 41 K. STM images presented were obtained with the tip bias at +1.5 V and feedback current at 3 nA. The FTs are shown as insets of the corresponding real space images. (d)-(f) show the perpendicular distortion by filtering out the ($\sqrt{3} \times \sqrt{3}$) symmetry from images (a)-(c), respectively. Ge substitutional defects are marked with red, green, or blue circles depending on their coordination relative to the three nonequivalent sites in a (3×3) lattice. Three or more neighboring defects are marked in light blue in (f). In (e), the (3×3) domains are highlighted in colors of the "up" sites. The adatoms distorted downwards are connected with gray lines in (f).

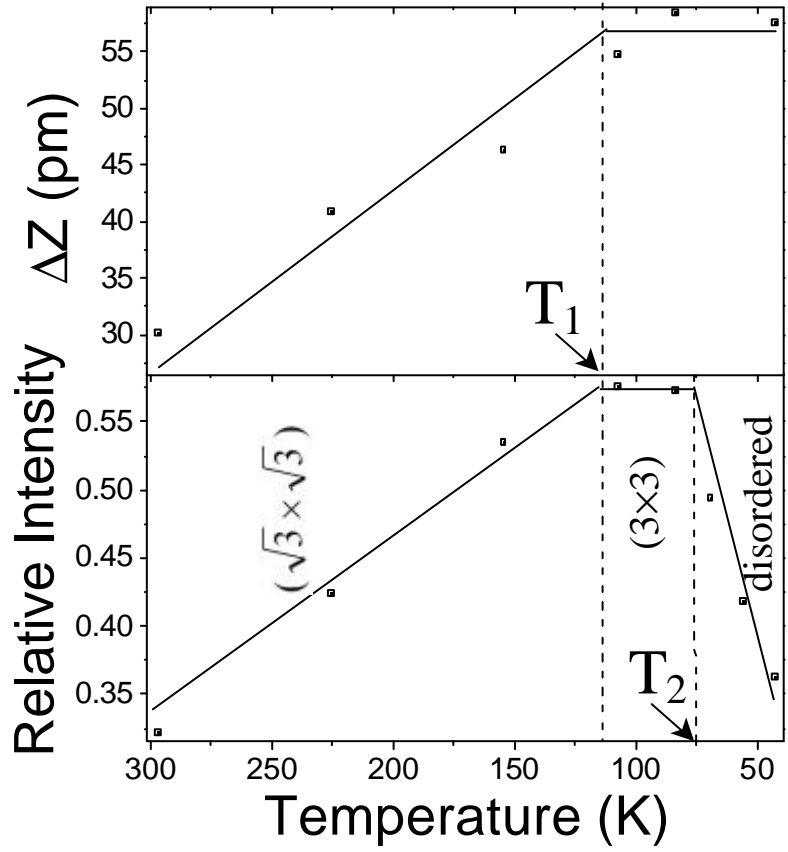

FIG. 2  Temperature dependence of the two order parameters of Pb/Ge interface $\Delta Z(T)$ and $I_3/I_{\sqrt{3}}(T)$. The height Z was obtained with the tip bias at +1.5 V and feedback current at 3 nA.

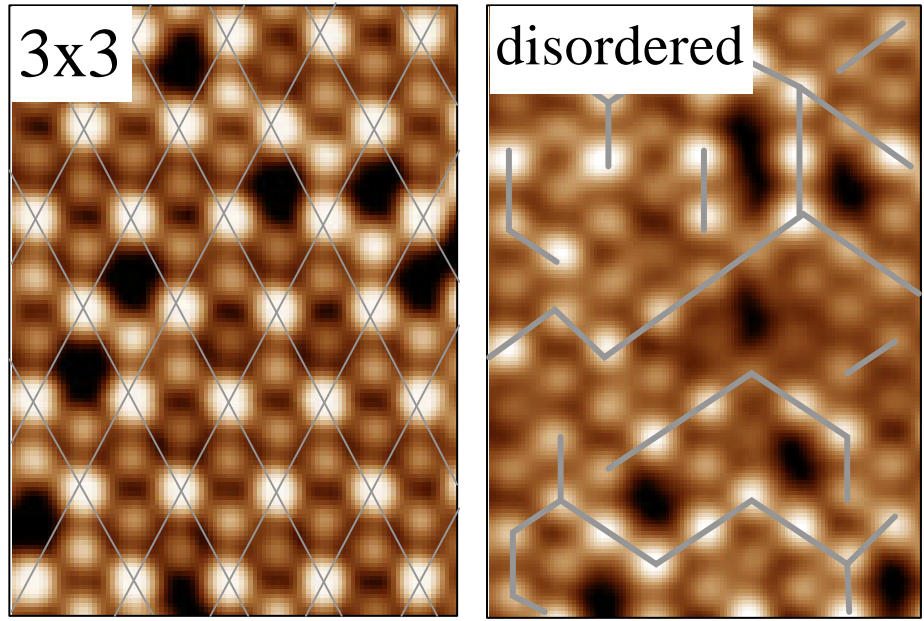

FIG. 3  STM images of a (3×3) single-domain at 90 K and an area in the disordered phase at 40 K with the same defect distribution with respect to the virtual (3×3) lattice. The gray grids in the left panel mark the charge-maximum sites of the (3×3) lattice. The kinked lines structure is indicated in the right panel by the gray lines connecting the adatoms distorted upwards.